\documentclass[12pt]{article}
\usepackage{amsmath,amssymb,bm}
\usepackage{graphicx}
\usepackage{enumerate}
\usepackage{natbib}

\usepackage[colorlinks=true,allcolors=blue]{hyperref}%
\usepackage{url} 
\usepackage[percent]{overpic}
\usepackage{caption}
\usepackage{graphicx,subfigure}
\usepackage{dsfont}
\usepackage{algorithm,algorithmic}
\usepackage[normalem]{ulem}
\usepackage{xcolor}
\usepackage{epstopdf,multirow,lscape,comment}
\usepackage{amsthm}

\newcommand{\blind}{0}

\newtheorem{Th}{{\bf Theorem}}

\newtheorem{proposition}{Proposition}

\newtheorem{Ass}{Assumption}
\newtheorem*{Ass*}{Assumption 1.1}

\def\A{{\bm A}}

\def\b{{\bm b}}

\def\1{{\bm 1}}
\def\0{{\bm 0}}

\def\mS{\mathcal{S}}

\def\mG{\mathcal{G}}

\def\bA{{\mathbf A}}
\def\ba{{\mathbf a}}

\def\bB{{\mathbf B}}

\def\bD{{\mathbf D}}

\def\bF{{\mathbf F}}
\def\b1e{{\mathbf e}}

\def\bI{{\mathbf I}}

\def\bB{{\mathbf B}}

\def\bR{{\mathbf R}}

\def\bS{{\mathbf S}}

\def\by{{\mathbf y}}
\def\bY{{\mathbf Y}}

\def\bz{{\mathbf z}}

\def\bx{{\mathbf x}}
\def\bX{{\mathbf X}}

\def\bzero{{\mathbf 0}}

\def\bepsilon{{\boldsymbol{ \epsilon}}}

\def\bbeta{{\boldsymbol{\beta}}}

\def\bphi{{\boldsymbol{\phi}}}
\def\bPhi{{\boldsymbol{\Phi}}}

\def\boxit#1{\vbox{\hrule\hbox{\vrule\kern6pt
          \vbox{\kern6pt#1\kern6pt}\kern6pt\vrule}\hrule}}

\def\wt{\widetilde}

\def\wh{\widehat}

\def\argmin{\hbox{argmin}}

\def\bse{\begin{eqnarray*}}
\def\ese{\end{eqnarray*}}
\def\be{\begin{eqnarray}}
\def\ee{\end{eqnarray}}
\def\bq{\begin{equation}}
\def\eq{\end{equation}}
\def\bse{\begin{eqnarray*}}
\def\ese{\end{eqnarray*}}

\def\wh{\widehat}

\def\b1e{{\mathbf e}}

\def\bx{{\mathbf x}}
\def\bX{{\mathbf X}}

\def\bS{{\mathbf S}}
\def\bT{{\mathbf T}}
\def\bzero{{\mathbf 0}}

\newcommand*{\mydot}{\mathrel{\scalebox{0.4}{$\bullet$}}}

\newcommand{\bgamma}{\mbox{\boldmath $\gamma$}}

\newcommand{\bSigma}{\mbox{\boldmath $\Sigma$}}

\newcommand{\rkc}[1]{{\color{red}[RK: #1]}}

\newcommand{\rkcr}[1]{{\color{red}#1}}
\newcommand \jz[1] {{\color{cyan}{[JZ: #1]}}}

\begin{document}

\if0\blind
{
  \title{\bf A covariate-dependent Cholesky decomposition for high-dimensional covariance regression}
  \author{Rakheon Kim\\
    Department of Statistical Science, Baylor University\\
    and \\
    Jingfei Zhang \\
    Goizueta Business School, Emory University}
    \date{}
  \maketitle
} \fi

\if1\blind
{
  \bigskip
  \bigskip
  \bigskip
  \begin{center}
    {\LARGE\bf Title}
\end{center}
  \medskip
} \fi

\bigskip
\begin{abstract}
\baselineskip=17.5pt
Estimation of covariance 
matrices is a fundamental problem in multivariate statistics.  
Recently, 
growing efforts have focused on incorporating covariate effects into these matrices, facilitating subject-specific estimation. Despite these advances, guaranteeing the positive definiteness of the resulting estimators remains a challenging problem.
In this paper, we present a new varying-coefficient sequential regression framework that extends the modified Cholesky decomposition to model the positive definite covariance matrix as a function of subject-level covariates. 
To handle high-dimensional responses and covariates, we impose a joint sparsity structure that simultaneously promotes sparsity in both the covariate effects and the entries in the Cholesky factors that are modulated by these covariates.
We approach parameter estimation with a blockwise coordinate descent algorithm, and investigate the $\ell_2$ convergence rate of the estimated parameters. 
The efficacy of the proposed method is demonstrated through numerical experiments and an application to a gene co-expression network study with brain cancer patients. 
\end{abstract}

\noindent%
\vfill

\newpage
\baselineskip=26.5pt

\section{Introduction}
\label{sec:intro}

Estimation of covariance matrices is fundamental for uncovering 
associations among variables and has been widely applied in genetics \citep{butte2000discovering, su2023cell}, neuroscience \citep{zhang2020mixed,zhang2023generalized}, finance \citep{el2010high, xue2012positive} and climotology \citep{bickel2008covariance}. In genetics, for instance, covariance estimated from gene expression across biological samples is used to identify functional gene modules and dysregulated pathways in disease \citep{langfelder2008wgcna,su2023cell}. Importantly, the structure and degree of such associations among genes may themselves vary with subject-level covariates (e.g. age, sex, genotype). 
A genetic variant affecting co-expression between two genes is known as a co-expression quantitative trait loci (QTL), and identifying such loci is crucial in developing gene therapies that target specific gene or pathway disruptions \citep{van2018integrative,zhang2023eqtl}.

Covariate-dependent covariance estimation has been addressed primarily in regression frameworks. \citet{chiu1996matrix} modeled elements in the logarithm of the covariance matrix as a linear function of covariates, though parameter interpretation was limited. Quadratic covariance regression \citep{hoff2012covariance, fox2015bayesian, franks2021reducing, alakus2022covariance} admits a nice random-effect model representation but is computationally intensive in high dimensions. Principal regression approaches \citep{zhao2021covariate, park2023bayesian, he2024covariance} also models the covariate effect on covariance but they avoid direct modeling of entries of the covariance matrix, complicating interpretation. \citet{zou2017covariance, zou2022inference} link the covariance matrix to similarity matrices of covariates but these methods typically assume such similarity matrices are known, which may not be available in real applications. 

Recently, \citet{kim2024high} proposed a covariance regression 
that models covariance as a linear function of subject-level covariates.
However, the resulting covariance matrix 
is not guaranteed to be positive definite (PD). 
\citet{he2025positive} used a similar covariance regression model as in \citet{kim2024high}, 
and proposed an alternating direction method of multipliers algorithm to guarantee positive definiteness. 
However, they focus on cases where repeated measurements of the response variables are available, and computational complexity of their proposed 
algorithm is exponential with respect to the number of covariates, making it infeasible with large number of covariates. Another line of work 
investigates Gaussian graphical models where the inverse covariance matrix is modeled as a linear function of covariates 
\citep{zhang2022high, zhang2022multi}. 
However, these methods also fail to guarantee positive definiteness 
of the inverse covariance matrix. In addition, the diagonal entries in the inverse covariance matrix are assumed to be independent of the covariates, which can be restrictive. 

Ensuring the positive definiteness of covariate matrix estimators has been extensively studied when they do not depend on covariates.  
\citet{ledoit2004well} introduced linear shrinkage 
to ensure positive definiteness, 
which was later extended to non-linear shrinkage \citep{ledoit2012nonlinear}.
Assuming sparsity in the covariance matrix, \citet{xue2012positive, rothman2012positive, wen2016positive, wen2021fast} proposed positive definite approximation for thresholding estimators and \citet{kim2023positive, fatima2024two} proposed maximum likelihood estimation to refit the non-zero elements in thresholding estimators.
\citet{bien2011sparse} proposed a majorization-minorization algorithm to compute a penalized likelihood estimator. 
\citet{Friedman2008} proposed the graphical lasso algorithm to compute a sparse inverse covariance matrix.
When the variables of interest have a natural ordering such as in time series or longitudinal data, \citet{pourahmadi1999joint} suggested an unconstrained parameterization of covariance matrices via the modified Cholesky decomposition, which ensures the covariance matrix estimator to be positive definite. The Cholesky factors from the 
decomposition can be interpreted as 
the generalized autoregressive parameters, encoding conditional dependence among variables in a sequential regression formulation, and the modified Cholesky decomposition has been extended by \citet{huang2006covariance, levina2008sparse} to high dimensional settings.

In this paper, we propose a new 
parameterization of covariance matrices which accounts for the effect of subject-level covariates via a Cholesky-based decomposition. This decomposition, which we call covariate-dependent Cholesky decomposition, provides closed-form estimates of both the covariate-dependent covariance matrix and inverse covariance matrix, whereas previous works on covariance matrix \citep{kim2024high} and inverse covariance matrix \citep{zhang2022high} do not produce closed-form solutions to both matrices. As the Cholesky factors in covariate-dependent Cholesky decomposition are unconstrained, our proposed method facilitates the positive definite estimation of subject-level covariance matrices which is not guaranteed in \citet{zhang2022high} and \citet{kim2024high}. 
Also, the Cholesky factors and the diagonal parameters in this decomposition have a nice statistical interpretation as the regression coefficients and the prediction error variance in a varying-coefficient linear model, respectively. 
Furthermore, compared to \citet{zhang2022high, zhang2022multi}, our proposed method does not impose any constraint on the diagonals of the inverse covariance matrix and does not require Gaussian assumption on the response variables.

For the unconstrained Cholesky factors, we propose a penalized regression for the varying-coefficient linear model with simultaneous sparsity which selects effective covariates and their effects on the covariance matrix. For the error variance, we adopt the log-link to respect the positive constraint of variance and propose a penalized regression which selects effective covariates.
Our method assumes there is a natural ordering among the response variables, which is available in many applications. In our motivating data example of gene network analysis, the ordering of the genes can be retrieved from reference signaling pathway such as the KEGG human glioma pathway \citep{kanehisa2000kegg}.
Although motivated by gene co-expression analysis, our method is broadly applicable to other scientific areas such as neuroscience, finance and psychology that involve covariate-dependent covariance or inverse covariance estimation \citep{lin2025covariate, liu2026time}. 

The rest of the paper is organized as follows. Section \ref{sec:meth} introduces the Covariate-Dependent Cholesky Decomposition and Section \ref{sec:est} discusses its estimation with sparsity. Section \ref{sec:theory} investigates theoretically the convergence rate of the proposed estimator. Section \ref{sec:simul} carries out comprehensive simulation studies and Section \ref{sec:real} conducts a co-expression QTL analysis using a brain cancer genomics data set. A short discussion section concludes the paper.

\section{Modified Cholesky Decomposition with Covariates}
\label{sec:meth}
We start with some notation. Write $[d]=\{1,2,\ldots,d\}$. Given a vector $\bx=(x_1, \ldots, x_d)^\top$, we use $\Vert\bx\Vert_1$, $\Vert\bx\Vert_2$ and $\Vert\bx\Vert_{\infty}$ to denote the vector $\ell_1$, $\ell_2$ and $\ell_{\infty}$ norms, respectively.
We use $\lambda_{\min}(\cdot)$ and $\lambda_{\max}(\cdot)$ to denote the smallest and largest eigenvalues of a matrix, respectively.
For a matrix $\bX\in\mathbb{R}^{d_1\times d_2}$, we let $\Vert\bX\Vert_1 = \sum_{ij}|X_{ij}|$, $\Vert\bX\Vert_F=(\sum_{ij}X_{ij}^2)^{1/2}$, $\Vert\bX\Vert_2=\sqrt{\lambda_{\max}(\bX^\top \bX)}$ and $\Vert\bX\Vert_{\infty}=\max_{ij}|X_{ij}|$ denote the matrix element-wise $\ell_1$ norm, Frobenius norm, operator norm and element-wise max norm, respectively, and let $\text{vech}(\bX)=(X_{11}, X_{12},X_{22},X_{13},\ldots,X_{1,d_1},\ldots,X_{d_1d_1})^\top$ represent the vectorization of the upper triangular part of $\bX$ and $\text{vec}(\bX)$ represent the concatenation of columns in $\bX$. 

Given a vector of $p$ ordered response variables denoted as $\by=(y_1,\ldots, y_p)^\top$, and a vector of $q$ covariates denoted as $\bx=(x_1,\ldots, x_q)^\top$, we assume that 
$$
\mathbb{E}(\by|\bx)=\bgamma_0+\bm\Gamma\bx,\quad \text{Cov}(\by|\bx)=\bSigma(\bx),
$$
where $\bgamma_0\in\mathbb{R}^{p}$, $\bm\Gamma\in\mathbb{R}^{p\times q}$.
To expose key ideas, we assume $\bgamma_0$ and $\bm\Gamma$ are known in the ensuing development and $\by$ is demeaned, so that we focus on the estimation of $\bSigma(\bx)$. Extensions with estimated $\bgamma_0$ and $\bm\Gamma$ are straightforward, but with more involved notation.

We model $\bSigma(\bx)$ by considering its unique decomposition as
\be \label{eq:covariateMCD}
\bT(\bx) \bSigma(\bx) \bT(\bx)^\top = \bD(\bx) \quad \text{and} \quad \bSigma^{-1}(\bx) = \bT(\bx)^\top \bD^{-1}(\bx) \bT(\bx),
\ee
where $\bT(\bx)$ is a lower triangular matrix with all diagonal entries equal to one and $\bD(\bx)$ is a diagonal matrix with non-negative entries. 
Since both $\bT(\bx)$ and $\bD(\bx)$ depend on $\bx$, we refer to this decomposition of $\bSigma(\bx)$ as the \textit{covariate-dependent Cholesky decomposition (CDCD)}. 
When there are no covariates, \eqref{eq:covariateMCD} reduces to the standard modified Cholesky decomposition of a covariance matrix $\bSigma$ by $\bT \bSigma \bT^\top = \bD$ \citep{pourahmadi1999joint}. 
By property of the modified Cholesky decomposition \citep{pourahmadi1999joint}, the elements of $\bT(\bx)$ have the statistical interpretation as the regression coefficients $f_{tj}(\bx)$ 
in the following recursive regressions 
\begin{align} \label{eq:varyingAR}
y_t &= \epsilon_t \quad\quad\quad\quad\quad\quad\quad\quad\quad\quad \text{if} \quad t=1 \nonumber \\
&= \sum_{j=1}^{t-1} f_{tj}(\bx) y_j + \epsilon_t, \quad\quad\quad \; \text{if} \quad t>1
\end{align}
and the diagonal entries of $\bD(\bx)$ represent the prediction error variance $\text{Var}(\epsilon_t)$.
This is because, denoting $\bepsilon=(\epsilon_1,\ldots,\epsilon_p)^\top$, the model \eqref{eq:varyingAR} can be written as $\bT(\bx) \by = \bepsilon$ where $\{\bT(\bx)\}_{tj}=-f_{tj}(\bx)$ for $j<t$ and 
\bse
\text{Cov}(\bepsilon|\bx) = \text{Cov}\{\bT(\bx) \by |\bx\} = \bT(\bx) \bSigma(\bx) \bT(\bx)^\top,
\ese
which is a diagonal matrix since the residuals from the regressions \eqref{eq:varyingAR} are uncorrelated. We denote $\bD(\bx)=\text{Cov}(\bepsilon|\bx)$ as a diagonal matrix with $\sigma_t^2(\bx)$ as its $t$th diagonal element.
By construction of the model \eqref{eq:varyingAR}, $\{-\bT(\bx)\}_{tj}$ represents the strength and direction of the linear dependence of $y_t$ on $y_j$ given the other preceding variables, $y_1,\ldots,y_{j-1},y_{j+1},\ldots,y_{t-1}$.

One advantage of the CDCD formulation is that it provides closed-form estimates of both the covariate-dependent covariance matrix and inverse
covariance matrix. To see this, define a lower triangular matrix $\bF(\bx)$ such that $\bT(\bx)=\bI - \bF(\bx)$. From \eqref{eq:covariateMCD} and by the Neumann series expansion \citep{horn2012matrix}, we have
\begin{align*}
\bSigma(\bx) &= \{\bT(\bx)\}^{-1} \bD(\bx) \{\bT(\bx)^\top\}^{-1}\\
&= \bigg\{\sum_{j=0}^{p-1} \bF(\bx)^j\bigg\} \bD(\bx) \bigg\{\sum_{j=0}^{p-1} \bF(\bx)^j\bigg\}^\top.
\end{align*}
Hence, the estimation of $\bT(\bx)$ and $\bD(\bx)$ by the CDCD gives the closed-form estimates not only for $\bSigma^{-1}(\bx)$ by \eqref{eq:covariateMCD} but also for $\bSigma(\bx)$.

For $f_{tj}(\bx)$, we consider the linear function of $\bx$ as below
\bse
f_{tj}(\bx) = \phi_{t,j,0} + \sum_{k=1}^{q} \phi_{t,j,k} x_k,
\ese
where $\phi_{t,j,k}$'s are unconstrained.
This $f_{tj}(\bx)$ turns the model \eqref{eq:varyingAR} into a linear varying coefficient model which is a special case of an interaction model \citep{Lim2015, She2018}. A similar model 
has been studied in high dimensional linear regression \citep{tibshirani2020pliable, kim2021svreg}
and in Gaussian graphical model \citep{zhang2022high}.
Correspondingly, $\bT(\bx)$ can be expressed in a matrix form as
\be \label{eq:modelT}
\bT(\bx) = \bT_0 + \sum_{k=1}^{q} x_k \bT_k,\quad\text{where}
\ee
\bse
\bT_k = 
\begin{bmatrix}
\mathds{1}_{\{k=0\}} & 0 & 0 & ... & 0 & 0 \\
-\phi_{2,1,k} & \mathds{1}_{\{k=0\}} & 0 & ... & 0 & 0 \\
-\phi_{3,1,k} & -\phi_{3,2,k} & \mathds{1}_{\{k=0\}} & ... & 0 & 0 \\
\vdots & \vdots & \vdots & \ddots & \vdots & \vdots \\
-\phi_{p-1,1,k} & -\phi_{p-1,2,k} & -\phi_{p-1,3,k} & ... & \mathds{1}_{\{k=0\}} & 0 \\
-\phi_{p,1,k} & -\phi_{p,2,k} & -\phi_{p,3,k} & ... & -\phi_{p,p-1,k} & \mathds{1}_{\{k=0\}} \\
\end{bmatrix}
\ese
and $\mathds{1}$ represents the indicator function.
As the contribution of $\bT_0$ to $\bT(\bx)$ does not depend on the covariates, the entries of $\bT_0$ represent the linear effects among response variables at the population level. That is, $\phi_{t,j,0}$ measures the linear association between $y_j$ and $y_t$ given the predecessors of $y_t$ at the population level. The association between $y_j$ and $y_t$ given the predecessors of $y_t$ may vary across individuals depending on the value of covariates and the influence of each covariate to the linear association as represented by $\phi_{t,j,k}, k \in [q]$.

Without any constraint on $\bT_k$,
$\bSigma(\bx)$ is guaranteed to be positive definite as long as the diagonals in $\bD(\bx)$ are positive. 
Diagonals in $\bD(\bx)$ can be estimated by modeling the logarithm of $\sigma_t^2(\bx), t \in [p]$, as in the variance regression model \citep{harvey1976estimating}
\be \label{eq:modelD}
\log \sigma_t^2(\bx) = \beta_{t,0} + \sum_{k=1}^q \beta_{t,k} x_k.
\ee

\section{Estimation}
\label{sec:est}

\subsection{Estimation of cholesky factors}

With $n$ independent observations denoted as $\{(\by_i,\bx_i), i\in[n]\}\in\mathbb{R}^p\times\mathbb{R}^q$, we aim to estimate $\bT_0,\bT_1,\ldots,\bT_q$ in \eqref{eq:modelT} without imposing distributional assumptions on $\by_i$'s.
When both $p$ and $q$ are large, to ensure the estimability and facilitate the interpretability, we impose $\bT_0,\bT_1,\ldots,\bT_q$ to be sparse. To achieve this, we estimate the regression coefficients in \eqref{eq:modelT} by minimizing the penalized least squares as
\be \label{eq:obj_ftn}
\frac{1}{2n}\sum_{i=1}^n\sum_{t=2}^p \bigg( y_{it} - \sum_{j=1}^{t-1} \sum_{k=0}^{q} \phi_{t,j,k} x_{ik} y_{ij} \bigg)^2 + \mathcal{P}(\bT_0,\bT_1,\ldots,\bT_q)
\ee
where $x_{ik}$ and $y_{ij}$ are the $i$th observations for $x_k$ and $y_j$, respectively, and $\mathcal{P}(\cdot)$ is a penalty function.
In particular, we assume $\bT_1,\ldots,\bT_q$ are \textit{group sparse} so that 
only a subset of the covariates may impact the covariance matrix (termed effective covariates). Since $\bT_0$ represents the covariance at population level and is not related to any covariate, it is not subject to the group sparsity.
We further assume each $\bT_k, k=0,1,\ldots,q$ is \textit{element-wise sparse}. That is, effective covariates may influence only a subset of elements in $\bT_k$'s. These simultaneous sparsity assumptions are  well supported by genetic studies \citep{gardner2003inferring, vierstra2020global}, and improve model interpretability when compared to using the group sparsity or element-wise sparsity alone.


We consider the following penalty
\be \label{eq:pen}
\mathcal{P}(\bT_0,\bT_1,\ldots,\bT_q)= \lambda \sum_{k=0}^q \sum_{t=2}^p \sum_{j=1}^{t-1} |\phi_{t,j,k}| + \lambda_g \sum_{k=1}^q \|\bphi_{\mydot,\mydot,k}\|_2,
\ee
where 
$\bphi_{\mydot,\mydot,k}$ is a vector of nonredundant elements in $\bT_k$ and $\lambda, \lambda_g$ are tuning parameters. 
The term $\sum_{k=0}^q \sum_{t=2}^p \sum_{j=1}^{t-1} |\phi_{t,j,k}|$ is a lasso penalty that encourages the effect of effective covariates to be sparse. 
The term $\sum_{k=1}^q \|\bphi_{\mydot,\mydot,k}\|_2$ is a group lasso penalty \citep{Yuan2006} that encourages the effective covariates to be sparse, achieved by regularizing $\bT_k$ across $p-1$ regression tasks from \eqref{eq:varyingAR} simultaneously. 
Correspondingly, this penalty term facilitates a multi-task learning approach \citep{argyriou2008convex}.
The penalty term in \eqref{eq:pen} is similar to the sparse group lasso considered in \citet{Simon2013, li2015multivariate}, though it is not exactly the same as some parameters are included in the element-wise sparsity penalty but not the group sparsity penalty. This adds additional complexity to the estimation procedure and theoretical analysis. 


Let $\bY=(y_{it})$ be an $n \times p$ matrix and $\bX=(x_{ik})$ be an $n \times q$ matrix for $n$ observations of $\by$ and $\bx$, respectively. 
Denoting $\bPhi_k, k \in \{0,1,\ldots,q\}$ as the $(p-1) \times (p-1)$ upper triangular matrices containing the negatives of lower off-diagonal elements in $\bT_k$ as below
\bse
\bPhi_k = 
\begin{bmatrix}
\phi_{2,1,k} & \phi_{3,1,k} & \phi_{4,1,k} & ... & \phi_{p,1,k} \\
0 & \phi_{3,2,k} & \phi_{4,2,k} & ... & \phi_{p,2,k} \\
0 & 0 & \phi_{4,3,k} & ... & \phi_{p,3,k} \\
\vdots & \vdots & \vdots & \ddots & \vdots \\
0 & 0 & 0 & ... & \phi_{p,p-1,k}
\end{bmatrix},
\ese
the objective function \eqref{eq:obj_ftn} can be written in matrix form as
\be \label{eq:obj_ftn2}
\frac{1}{2n} \|\bY_{\mydot,2:p} - \sum_{k=0}^q (\bY_{\mydot,1:(p-1)}\circ\bX_{\mydot,k})\bPhi_k\|_F^2 + \lambda \sum_{k=0}^q \|\bPhi_k\|_1 + \lambda_g \sum_{k=1}^q \|\bPhi_k\|_F,
\ee
where $\bY_{\mydot,a:b}, a<b$ is a matrix containing from the $a$th column to the $b$th column in $\bY$, $\bX_{\mydot,k}$ is the $k$th column in $\bX$ for $k \in [q]$, $\bX_{\mydot,0}$ is defined as an $n \times 1$ matrix of ones and $\circ$ denotes element-wise multiplication.
Hence, $\bY_{\mydot,1:(p-1)}\circ\bX_{\mydot,k}$ terms for $k \in [q]$ represent the interaction terms of $y_j, j \in [p-1]$ and $x_k$ in the model \eqref{eq:varyingAR}.

For optimization of \eqref{eq:obj_ftn2}, we adopt the blockwise coordinate descent algorithm as described in Algorithm \ref{alg1}. 
For $k=0$, the solution to $\bPhi_k$ is obtained by the lasso estimator, as the intercept terms in \eqref{eq:varyingAR} are not penalized by the group lasso penalty. For $k \in [q]$, the solution to $\bPhi_k$ is obtained by the sparse group lasso estimator. 
Define $\bX_{\mydot,0}$ as an $n \times 1$ matrix of ones. 
Specifically, the Karush-Kuhn-Tucker condition for the sparse group lasso \citep{Simon2013} 
is satisfied with $\bPhi_k=\bzero$ if
\be \label{eq:zero_cond}
\|\text{vech}[ S_\lambda \{(\bY_{\mydot,1:(p-1)}\circ\bX_{\mydot,k})^\top \bR_k/n \} ]\|_2 \leq \lambda_g,
\ee
where $\bR_k=\bY_{\mydot,2:p} - \sum_{l \neq k} (\bY_{\mydot,1:(p-1)}\circ\bX_{\mydot,l})\bPhi_l$ is the partial residual of $\bY_{\mydot,2:p}$ 
and $S_{\lambda}(\bA)$ is the element-wise soft-thresholding operator at $\lambda$ for a matrix $\bA=(a_{ij})$, that is, $\{S_{\lambda}(\bA)\}_{ij} = \text{sign}(a_{ij})\times\max(|a_{ij}|-\lambda, 0)$. 
When $\bPhi_k \neq \bzero$, the solution to $\phi_{t,j,k}$ for  $t \in \{2,\ldots,p\}$ and $j \in [t-1]$ is given by
\be \label{eq:nonzero_sol}
\hat{\phi}_{t,j,k} = \frac{S_\lambda\{(\bY_{\mydot,j}\circ\bX_{\mydot,k})^\top \bR_{t,j,k}/n\}}{ \|\bY_{\mydot,j}\circ\bX_{\mydot,k}\|_2^2/n + \mathds{1}_{\{k \neq 0\}} \lambda_g / \|\bPhi_k\|_F },
\ee
where 
$
\bR_{t,j,k} = \bY_{\mydot,t} - \sum_{l \neq k} (\bY_{\mydot,1:(p-1)}\circ\bX_{\mydot,l})(\bPhi_l)_{\mydot,t-1} - \sum_{\substack{m \neq j \\ m \in [t-1]}} (\bY_{\mydot,m}\circ\bX_{\mydot,k})\phi_{t,m,k}.
$
Detailed derivation of the optimization is described in Supplementary Materials S1. 
Two parameters $\lambda$ and $\lambda_g$ in \eqref{eq:obj_ftn} require tuning. In our procedure, they are jointly selected via $L$-fold cross validation. 

\begin{algorithm}[!t]
\caption{Covariate-dependent Cholesky decomposition}\label{alg1}
\begin{algorithmic}
\STATE \textbf{Input:} Tuning parameters $\lambda$, $\lambda_g$, and initial estimators $\wt{\bPhi}_k$ 
for $k=0,1,\ldots,q$.
\vspace{0.05in}
\STATE \textbf{Iterate} over $k=0,1,\ldots,q$ 
\STATE \hspace{0.25in} \textbf{Step 1:} For $j=1,\ldots,t-1$ and $t=2,\ldots,p$, compute $\wt{\bR}_{t,j,k}$ as:
$$
\wt{\bR}_{t,j,k} = \bY_{\mydot,t} - \sum_{\substack{l \neq k \\ l \in \{0\}\cup[q] }} (\bY_{\mydot,1:(p-1)}\circ\bX_{\mydot,l})(\wt{\bPhi}_l)_{\mydot,t-1} - \sum_{\substack{m \neq j \\ m \in [t-1]}} (\bY_{\mydot,m}\circ\bX_{\mydot,k})\tilde{\phi}_{t,m,k}.
$$
\STATE \hspace{0.25in} \textbf{Step 2:} For $k=0$, update $\tilde{\phi}_{t,j,0}$ for $j=1,\ldots,t-1$ and  $t=2,\ldots,p$ by
$$
\tilde{\phi}_{t,j,0} = \bigg( \frac{1}{n} \| \bY_{\mydot,j} \|_2^2 \bigg)^{-1} S_{\lambda}\bigg(\frac{1}{n} \bY_{\mydot,j}^\top \wt{\bR}_{t,j,0} \bigg).
$$
\STATE \hspace{0.25in} \textbf{Step 3:} For $k\neq 0$, compute $\bR_{k}$ as:
$
\wt{\bR}_{k}=\bY_{\mydot,2:p} - \sum_{\substack{l \neq k \\ l \in \{0,1,\ldots,q\} }} (\bY_{\mydot,1:(p-1)}\circ\bX_{\mydot,l})\wt{\bPhi}_l
$
\STATE \hspace{0.9in} and check the condition
                $
			    \bigg\| \text{vech}\bigg[ S_{\lambda}\bigg\{\frac{1}{n} (\bY_{\mydot,1:(p-1)}\circ\bX_{\mydot,k})^\top \widetilde{\bR}_{k} \bigg\} \bigg] \bigg\|_2 < \lambda_g$.
\STATE \hspace{0.8in} $\rhd$ If the condition above is satisfied, set $\wt{\bPhi}_k=\bzero$.
\STATE \hspace{0.8in} $\rhd$ If not, update $\tilde{\phi}_{t,j,k}$ for for $t=2,\ldots,p$ and $j=1,\ldots,t-1$ by
\hspace{0.1in} 
$$
\tilde{\phi}_{t,j,k} = \bigg( \frac{1}{n} \| \bY_{\mydot,j}\circ\bX_{\mydot,k} \|_2^2 + \frac{\lambda_g}{\|\wt{\bPhi}_k\|_F} \bigg)^{-1} S_{\lambda}\bigg\{\frac{1}{n} (\bY_{\mydot,j}\circ\bX_{\mydot,k})^\top \wt{\bR}_{t,j,k} \bigg\}.
$$
\STATE \textbf{until} the algorithm converges and obtain $\wh{\bT}_k$ by setting $\wh{\bPhi}_k = \wt{\bPhi}_k$ for $k=0,1,\ldots,q$.
\vspace{0.05in}
\STATE \textbf{Step 4:} Compute $\hat{\epsilon}_{it}$ by \eqref{eq:residuals} and fit \eqref{eq:modelD} via optimization of \eqref{eq:obj_ftn_D}.
\end{algorithmic}
\end{algorithm}

\subsection{Estimation of error variance}

Next, we consider the estimation of 
$\bbeta = \{\beta_{t,k}\}_{t=1,k=0}^{p,q}$ in \eqref{eq:modelD}. Due to the logarithmic operator in the model \eqref{eq:modelD}, estimation of the regression coefficients requires the use of the logarithmic link function as in the generalized linear model (GLM). 
Also, similar to the estimation of $\bT_k$'s, we assume group sparsity in the effective covariates, 
which requires the use of the group lasso \citep{Yuan2006}. 
Hence, we propose to estimate $\bbeta$ using
\be \label{eq:obj_ftn_D}
\wh{\bbeta} = \argmin_\bbeta \frac{1}{2n}\sum_{i=1}^n\sum_{t=1}^p \bigg( \hat{\epsilon}_{it}^2 - e^{ \beta_{t,0} + \sum_{k=1}^q \beta_{t,k} x_{ik} } \bigg)^2 + \lambda_d \sum_{k=1}^q \|\bbeta_{\mydot,k}\|_2
\ee
where $\hat{\epsilon}_{it}$ is computed from the fitted coefficients \eqref{eq:nonzero_sol} by
\begin{align} \label{eq:residuals}
\hat{\epsilon}_{it} &= y_{i1} \quad\quad\quad\quad\quad\quad\quad\quad\quad\quad\quad \text{if} \quad t=1 \nonumber \\
&= y_{it} - \sum_{j=1}^{t-1} \sum_{k=0}^{q} \hat{\phi}_{t,j,k} x_{ik} y_{ij} \quad \quad \; \text{if} \quad t \neq 1
\end{align}
and $\lambda_d$ is a tuning parameter for the group lasso penalty $\sum_{k=1}^q \|\bbeta_{\mydot,k}\|_2$.
The solution to \eqref{eq:obj_ftn_D} can be obtained by optimizing the quadratic approximation of the objective function \citep{meier2008group, friedman2010regularization}. Specifically, we adopt the blockwise coordinate descent which cycles through $k=0,1,\ldots,q$ by updating $\beta_{t,k}$ simultaneously for all $t \in [p]$ via the majorization-minorization as below 
\begin{align*}
\hat{\beta}_{t,k} & = \tilde{\beta}_{t,k} - g_{t,k}/h_{t,k}^\ast \quad \quad\quad\quad\quad\quad\quad\quad\quad\quad\quad\quad\quad\quad\quad\quad\quad \text{if} \quad k=0\\
&= \max \bigg(1-\frac{\lambda_d}{\|\tilde{\beta}_{t,k} - g_{t,k}/h_{t,k}^\ast\|_2}, 0 \bigg) (\tilde{\beta}_{t,k} - g_{t,k}/h_{t,k}^\ast) \quad\quad\; \text{if} \quad k>0
\end{align*}
where $\tilde{\beta}_{t,k}$ is the current value of $\beta_{t,k}$ for all $t \in [p], k \in \{0,1,\ldots,q\}$ and $g_{t,k}$ and  $h_{t,k}^\ast$ are
\begin{align*}
g_{t,k} &= \frac{1}{n}\sum_{i=1}^n \bigg( e^{ \tilde{\beta}_{t,0} + \sum_{k=1}^q \tilde{\beta}_{t,k} x_{ik} } - \hat{\epsilon}_{it}^2 \bigg) \bigg( e^{ \tilde{\beta}_{t,0} + \sum_{k=1}^q \tilde{\beta}_{t,k} x_{ik} } \bigg) x_{ik},\\
h_{t,k}^\ast &= \frac{1}{n}\sum_{i=1}^n 2\bigg( e^{ \tilde{\beta}_{t,0} + \sum_{k=1}^q \tilde{\beta}_{t,k} x_{ik} } \bigg)^2 x_{ik}^2.
\end{align*}
Here, the use of $h_{t,k}^\ast$ gives the surrogate function which majorizes the objective function \eqref{eq:obj_ftn_D}.
Detailed derivation of the optimization is given in Supplementary Materials S2. 

\section{Theoretical Properties}
\label{sec:theory}


We establish the non-asymptotic $\ell_2$ error rate of our proposed estimator from \eqref{eq:obj_ftn2} under the sub-Gaussian error assumption. 
One main challenge is that the error terms from the $p$ regression tasks \eqref{eq:varyingAR} are heterogeneous and heteroskedastic. 
Also, because the design matrix in \eqref{eq:obj_ftn2} includes high-dimensional interactions between $\by_i$ and $\bx_i$, and the variance of $\by_i$ is a function of $\bx_i$, characterizing their joint distribution is difficult. 
Lastly, as the combined penalty term $\lambda \sum_{k=0}^q \|\bPhi_k\|_1 + \lambda_g \sum_{k=1}^q \|\bPhi_k\|_F$ is not decomposable, the classic techniques for decomposable regularizers and null space \citep{negahban2012unified} are not applicable. 
By utilizing a tail bound for the quadratic form of sub-Gaussian variables, we derive a sharp bound for the stochastic term, yielding that our proposed estimator can have an improved $\ell_2$ error bounds compared to the lasso and the group lasso when the true coefficients are simultaneously sparse.


Let $\bphi=\{\phi_{t,j,k}\}_{t=2,j=1,k=0}^{p,t-1,q}$ be the vector of all non-redundant elements in $\bPhi_0,\ldots,\bPhi_q$.
Let $\mS$ be the element-wise support set of $\bphi$ and $\mG$ be the group-wise support set of $\bphi$ such that $\mG=\{k:\bPhi_k\neq \0, k\in[q]\}$, and denote by $s=|\mS|$ and $s_{g}=|\mG|$, that is, $s$ and $s_g$ are the numbers of nonzero entries and  nonzero groups, respectively. 
We state the needed regularity conditions. 
\begin{Ass}
\label{ass0}
Suppose $\epsilon_{it}$ is sub-Gaussian with mean zero and variance $\sigma_t^2(\bx_i)$ such that $\sigma_t^2(\bx_i) < M_0$ for $i\in[n], t\in[p]$. Suppose $\epsilon_{it}$ is independent of $\epsilon_{ih}$ for $h<t$. 
\end{Ass}

\begin{Ass}
\label{ass1}
Suppose $\bx_i$'s for $i \in [n]$ are independently and identically distributed mean zero sub-Gaussian random vectors with 
$\kappa_0^{-1} \le \lambda_{\min}(\text{Cov}(\bx_i)) \le \lambda_{\max}(\text{Cov}(\bx_i)) \le \kappa_0$ for a constant $\kappa_0>0$. 
\end{Ass}

\begin{Ass}
\label{ass2}
Suppose $\kappa_1^{-1}\le\lambda_{min}(\text{Cov}(\by_i))\le\lambda_{max}(\text{Cov}(\by_i))\le \kappa_1$ for a constant $\kappa_1>0$.
\end{Ass}
\begin{Ass}
\label{ass3}
The dimensions $p,q$ and sparsity $s$ satisfy $\log\,p+\log\,q=\mathcal{O}(n^\delta)$ and $s=o(n^\delta)$ for $\delta\in[0,1/6]$. 
\end{Ass}

Assumption \ref{ass0} on independent sub-Gaussian errors is applicable to Gaussian errors 
as their zero correlation assured by the recursive regression \eqref{eq:varyingAR} implies independence.
This assumption is also reasonable in many other settings, especially when there is a natural ordering such as in time series \citep{wu2012covariance}, longitudinal data analysis \citep{rhee2022robust}, Bayesian network \citep{loh2014high} and synthetic data generation \citep{kim2025synthetic}. 
Assumption \ref{ass0} also specifies that the variance of the errors in \eqref{eq:varyingAR} are bounded, which has been commonly adopted in other literature \citep{wu2003nonparametric}.
Assumptions \ref{ass1} and \ref{ass2} impose bounded eigenvalues on $\text{Cov}(\bx_i)$ and $\text{Cov}(\by_i)$ as commonly assumed in the high-dimensional regression literature \citep{chen2016asymptotically,cai2022sparse}. 
Assumption \ref{ass3} is a sparsity condition.

\begin{proposition} \label{lemma6}
Let $\bepsilon=(\epsilon_1,\ldots,\epsilon_n)$ be a vector of independent centered sub-Gaussian random variables with $K=\max_{i}\|\epsilon_i\|_{\psi_2}$. Let $\bSigma$ be the covariance matrix of $\bepsilon$ and $\bA$ be an idempotent matrix.
For any $t>0$, it holds that
$$
P\left(\|\bA \bepsilon\|_2^2\ge t\right)\le \exp\left[-\frac{\left\{t-tr(\A \bSigma)\right\}^2}{ c^2K^4\|\A\|_F^2 + 2cK^2\|\A\|_2 \left\{t-tr(\A \bSigma)\right\}}\right].
$$
\end{proposition}
Proposition \ref{lemma6} gives a tail bound for the quadratic form of a sub-Gaussian random vector. As the elements in $\bepsilon$ are allowed to have unequal variances, this bound allows us to address the heterogeneous errors in our model \eqref{eq:varyingAR}. 

Next, we discuss the $\ell_2$ convergence rate of our proposed estimator. Rewrite $\lambda=\alpha \lambda_0$ and $\lambda_g=(1-\alpha) \lambda_0$.
Let candidate models be those evaluated during tuning, and define $s_{\lambda}$ as the maximum number of nonzero coefficients across them such that $s<s_{\lambda}\le n$. 
In parameter tuning, given an $s_{\lambda}$ satisfying the conditions in Theorem \ref{thm2}, we choose the range of $\lambda_0$ for each $\alpha$, respectively corresponding to an empty model with no variables selected and a sparse model with $s_{\lambda}$ variables selected.

\begin{Th}\label{thm2}
Suppose that Assumptions \ref{ass0}-\ref{ass3} hold, $K=\max_{it}\|\epsilon_{it}\|_{\psi_2}$, $s_{\lambda}(\log\,p+\log\,q)=\mathcal{O}(\sqrt{n})$ and $n\ge A_1\{s_g\log(eq/s_g)+s\log(ep)\}$ for some constant $A_1>0$.
Then the estimator $\hat\bphi$ for \eqref{eq:obj_ftn2} with 
\begin{equation}
\lambda=C\sqrt{\log(eq/s_g)/n+2s\log(ep)/(ns_g)},\quad \lambda_g=\sqrt{s_g/s}\lambda
\end{equation}
satisfies, with probability at least $1-C_1\exp[-C_2\{s_g\log(eq/s_g)+s\log(ep)\}]$, 
\begin{equation}
\|\hat\bphi-\bphi\|_2^2\precsim \frac{1}{n}\{s_g\log(eq/s_g)+s\log(ep)\},
\end{equation}
where $C$, $C_1$, and $C_2$ are positive constants.
\end{Th}
Theorem \ref{thm2} shows that our proposed estimator enjoys an $\ell_2$ convergence rate that scales with both $s_g$ and $s$. 
The first term reflects the complexity of identifying the effective covariates, and the second term reflects the complexity of identifying the $s$ nonzero entries in the Cholesky factors.
It improves both over the standard lasso when $\log p/\log q=o(1)$ and over the group lasso when $\log q/\{p(p-1)\}=o(1)$ and $s/ s_g=o(p(p-1)/\log p)$, highlighting the benefit of using a combined regularizer. 
In Theorem \ref{thm2}, the condition $s_{\lambda}(\log p + \log q) = \mathcal{O}(\sqrt{n})$ limits model complexity and ensures control of the stochastic term. One may select $s_{\lambda}$ that is on the order of $c\sqrt{n}/\max(\log p, \log q)$ for some $c>0$, which implies $s = o(s_{\lambda})$ under Assumption \ref{ass3}.

\section{Simulation Studies}
\label{sec:simul}

In this section, we investigate the finite sample performance of our proposed method, referred to as \texttt{CDCD}, and compare it with other alternative methods, including:
\begin{itemize}
\item \texttt{DenseSample}: standard sample covariance estimator $\bS = \sum_{i=1}^n \bz_i \bz_i^\top /n$,
\vspace{-0.1in}
\item \texttt{SparseSample}: soft-thresholding sample covariance estimator $S_\lambda(\bS)$ where $S_\lambda(\cdot)$ is the element-wise soft-thresholding operator at $\lambda$ \citep{rothman2009generalized},
\vspace{-0.1in}
\item \texttt{SparseCovReg}: sparse covariance regression estimator in \citet{kim2024high}.
\end{itemize}
The tuning parameters for all methods are selected using 5-fold cross validation.


We simulate $n$ $(n=100,200)$ samples of $\{(\by_i,\bx_i), i\in[n]\}$, where the response $\by_i$ is of dimension $p$ (e.g., genes, $p=50$) and covariate $\bx_i$ is of dimension $q$ (e.g., genetic variants, $q=30,100$). 
For $\bx_i$'s, we consider 
binary covariates drawn independently from $\text{Bernoulli}(0.5)$.
Given $\bx_i$, we simulate $\by_i$ from $\mathcal{N}_{p}(\bzero, \bSigma(\bx_i))$, where the covariance structure follows one of the three models below. 
\begin{itemize}
\item First-order autoregressive, AR(1): $\{\Sigma(\bx)\}_{jk}=
\begin{cases}
    1,       & \quad \text{if } j=k,\\
    0.5^{|j-k|} x_1,  & \quad \text{if } j \neq k.
\end{cases}$
\item Hub graph: For $m \in \{0,1,2,3,4\}$ and $j \geq k$,\\
$\{\Sigma(\bx)^{-1}\}_{jk}=
\begin{cases}
    0.5,       &  \text{if } j=k \; \text{and} \; k\neq10m+1\\
    0.5 + 4.5 x_1 ,       &  \text{if } j=k \; \text{and} \; k=10m+1\\
    -0.5 x_1,  & \text{if } j \in \{10m+2,\ldots,10m+10\} \; \text{and} \; k=10m+1,\\
    0 ,       & \text{otherwise}. 
\end{cases}$
\item Random graph: 
$\bD(\bx)=\bI$ and $\bT(\bx)=\bI + x_1 \bT_1$ where 5\% of randomly chosen entries from the lower triangle of $\bT_1$ are equal to $-0.5$ and all other entries in $\bT_1$ are equal to zero. 
\end{itemize}
For the AR(1) model, the subjects with $x_1=1$ have the first-order autoregressive covariance structure which has zero values in all entries of the inverse covariance matrix except for the diagonal and the first off-diagonal entries. 
For the Hub graph model, the subjects with $x_1=1$ have the hub structure in the inverse covariance matrix and the block diagonal structure in the covariance matrix.
For the Random graph model, the subjects with $x_1=1$ have the inverse covariance matrix with randomly chosen non-zero entries.
These covariance structures have been commonly considered by others \citep{rothman2009generalized, bien2011sparse, qiu2019threshold, xu2022proximal}. 
For each simulation configuration, we generate 20 independent data sets.

\begin{table}[!t]
\centering
\scalebox{0.75}{
{\renewcommand{\arraystretch}{0.75}
\color{black}
\begin{tabular}{ccccccccc}
  \hline \hline
   &  &  & \multicolumn{2}{c}{\uline{AR(1)}} & \multicolumn{2}{c}{\uline{Hub}} & \multicolumn{2}{c}{\uline{Random}} \\[1.0ex]
   $n$ & $q$ & method & $\wh{\bSigma}_i$ & $\wh{\bSigma}_i^{-1}$ & $\wh{\bSigma}_i$ & $\wh{\bSigma}_i^{-1}$ & $\wh{\bSigma}_i$ & $\wh{\bSigma}_i^{-1}$ \\[0.5ex] \hline
   100 & 30 & \texttt{DenseSample} & 5.74 (0.14) & 19.70 (1.79) & 26.82 (0.80) & 10.01 (0.43) & 8.10 (0.72) & 16.77 (1.75) \\ [1.0ex]
     &  & \texttt{SparseSample} & 3.38 (0.17) & 4.52 (0.27) & 23.67 (0.90) & 6.02 (0.44) & 6.00 (0.93) & 4.11 (0.33) \\ [1.0ex]
     &  & \texttt{SparseCovReg} & 3.32 (0.17) & 4.42 (0.25) & 18.57 (1.25) & 5.84 (0.47) & 5.09 (0.63) & 4.10 (1.06) \\ [1.0ex]
     &  & \texttt{SparseCovReg(PD)} & 3.32 (0.17) & 4.42 (0.25) & 18.61 (1.24) & 5.83 (0.47) & 5.09 (0.63) & 3.87 (0.31)  \\ [1.0ex]
     &  & \texttt{CDCD} & \textbf{2.98} (0.13) & \textbf{3.94} (0.18) & \textbf{15.13} (0.95) & \textbf{4.04} (0.24) & \textbf{4.83} (0.63) & \textbf{3.35} (0.21)  \\ [1.0ex] \hline
   200 & 30 & \texttt{DenseSample} & 4.55 (0.08) & 7.90 (0.20) & 24.85 (0.47) & 6.74 (0.19) & 6.84 (0.75) & 6.89 (0.31) \\ [1.0ex]
     &  & \texttt{SparseSample} & 3.23 (0.10) & 4.42 (0.16) & 23.17 (0.52) & 5.93 (0.31) & 5.77 (0.92) & 3.95 (0.31) \\ [1.0ex]
     &  & \texttt{SparseCovReg} & 2.82 (0.08) & 3.89 (0.14) & 13.33 (1.30) & 5.45 (0.35) & 4.19 (0.59) & 3.42 (0.33) \\ [1.0ex]
     &  & \texttt{SparseCovReg(PD)} & 2.82 (0.08) & 3.89 (0.14) & 13.98 (1.61) & 5.47 (0.34) & 4.20 (0.61) & 3.40 (0.31)  \\ [1.0ex]
     &  & \texttt{CDCD} & \textbf{2.36} (0.08) & \textbf{3.09} (0.14) & \textbf{11.05} (0.97) & \textbf{3.15} (0.19) & \textbf{3.89} (0.60) & \textbf{2.58} (0.18)  \\ [1.0ex] \hline
    100 & 100 & \texttt{DenseSample} & 5.76 (0.12) & 19.29 (1.98) & 26.81 (0.67) & 10.23 (0.63) & 8.03 (0.69) & 17.29 (1.51) \\ [1.0ex]
     &  & \texttt{SparseSample} & 3.36 (0.19) & 4.50 (0.28) & 23.67 (0.81) & 6.04 (0.42) & 5.95 (0.96) & 4.10 (0.36) \\ [1.0ex]
     &  & \texttt{SparseCovReg} & 3.31 (0.17) & 4.42 (0.26) & 18.85 (1.22) & 5.93 (0.45) & 5.17 (0.75) & 3.91 (0.35) \\ [1.0ex]
     &  & \texttt{SparseCovReg(PD)} & 3.31 (0.17) & 4.42 (0.26) & 19.12 (1.56) & 5.88 (0.46) & 5.16 (0.75) & 3.90 (0.36)  \\ [1.0ex]
     &  & \texttt{CDCD} & \textbf{2.97} (0.11) & \textbf{3.98} (0.13) & \textbf{15.10} (0.91) & \textbf{4.23} (0.27) & \textbf{4.86} (0.71) & \textbf{3.34} (0.22) \\ [1.0ex] \hline
     200 & 100 & \texttt{DenseSample} & 4.54 (0.08) & 7.91 (0.37) & 24.65 (0.27) & 6.73 (0.19) & 6.80 (0.74) & 6.76 (0.27) \\ [1.0ex]
     &  & \texttt{SparseSample} & 3.23 (0.10) & 4.44 (0.16) & 22.96 (0.50) & 5.89 (0.29) & 5.73 (0.92) & 3.92 (0.28) \\ [1.0ex]
     &  & \texttt{SparseCovReg} & 2.79 (0.14) & 3.88 (0.19) & 14.26 (1.64) & 5.51 (0.33) & 4.36 (0.64) & 3.93 (1.54) \\ [1.0ex]
     &  & \texttt{SparseCovReg(PD)} & 2.79 (0.14) & 3.88 (0.19) & 14.26 (1.64) & 5.51 (0.33) & 4.39 (0.70) & 3.50 (0.34) \\ [1.0ex]
     &  & \texttt{CDCD} & \textbf{2.35} (0.12) & \textbf{3.10} (0.14) & \textbf{11.64} (1.01) & \textbf{3.29} (0.15) & \textbf{4.03} (0.64) & \textbf{2.65} (0.18) \\ [1.0ex]
  \hline \hline
\end{tabular}}}
\caption{Average error for individual covariance matrix $\wh{\bSigma}_i$ measured by $n^{-1} \sum_{i=1}^n \|\wh{\bSigma}_i-\bSigma_i^\ast\|_F$ and inverse covariance matrix $\wh{\bSigma}_i^{-1}$ measured by $n^{-1} \sum_{i=1}^n \|\wh{\bSigma}_i^{-1}-{\bSigma_i^\ast}^{-1}\|_F$ over 20 simulations with standard error shown in parentheses. The lowest error in each setting has been bolded.} 
\label{tb:frob}
\end{table}

Let $\bSigma_i^\ast$ denotes the true covariance matrix for the $i$th observation and $\wh{\bSigma}_i$ denotes the estimated $\bSigma_i^\ast$ from a given method.
We compare the average error for all individuals' covariance matrices measured by $n^{-1} \sum_{i=1}^n \|\wh{\bSigma}_i-\bSigma_i^\ast\|_F$.
We also compare for inverse covariance matrices measured by $n^{-1} \sum_{i=1}^n \|\wh{\bSigma}_i^{-1}-{\bSigma_i^\ast}^{-1}\|_F$.
Table \ref{tb:frob} reports the average errors with standard errors in the parentheses. 
The proposed \texttt{CDCD} outperforms the alternative methods for all $n$ and $q$.
Also, it is seen that the error of \texttt{CDCD} decreases as $n$ increases. 
Notably, by construction, \texttt{CDCD} always produces positive definite estimators of $\bSigma_i$. On the other hand, \texttt{SparseCovReg} does not guarantee the positive definiteness and \citet{kim2024high} proposed a post-hoc adjustment to satisfy a sufficient condition for positive definiteness estimation. 
For the Hub graph and Random graph models, \texttt{SparseCovReg} does not satisfy the condition for some simulated datasets, which results in different estimator of \texttt{SparseCovReg(PD)}, as shown by discrepancy in the average error in Table \ref{tb:frob}. 

\begin{table}[!t]
\centering
\scalebox{0.9}{
{\renewcommand{\arraystretch}{0.75}
\color{black}
\begin{tabular}{cccccc}
  \hline \hline \\ [-1.5ex] 
  $n$ & $q$ & Performance measure & \uline{AR(1)} & \uline{Hub} & \uline{Random} \\[1.0ex] \hline
  100 & 30 & $\|\hat\bphi-\bphi\|_2^2$ & 2.7865 & 3.7428 & 3.0690 \\ [1.0ex]
     &  & TPR & 0.8806 & 0.9967 & 0.8819 \\ [1.0ex]
     &  & FPR & 0.0142 & 0.0180 & 0.0153 \\ [1.0ex] \hline 
    200 & 30 & $\|\hat\bphi-\bphi\|_2^2$ & 2.0811 & 2.5872 & 2.2952 \\ [1.0ex]
     &  & TPR & 0.9980 & 1.0000 & 0.9899 \\ [1.0ex]
     &  & FPR & 0.0115 & 0.0098 & 0.0143 \\ [1.0ex] \hline 
     100 & 100 & $\|\hat\bphi-\bphi\|_2^2$ & 2.8165 & 3.9208 & 3.1052 \\ [1.0ex]
     &  & TPR & 0.9112 & 0.9978 & 0.8886 \\ [1.0ex]
     &  & FPR & 0.0072 & 0.0073 & 0.0046 \\ [1.0ex] \hline 
     200 & 100 & $\|\hat\bphi-\bphi\|_2^2$ & 2.0789 & 2.7792 & 2.3984 \\ [1.0ex]
     &  & TPR & 0.9949 & 1.0000 & 0.9916 \\ [1.0ex]
     &  & FPR & 0.0041 & 0.0037 & 0.0037 \\ [1.0ex] \hline 
  \hline \hline
\end{tabular}}}
\caption{Average estimation error measured by $\|\hat\bphi-\bphi\|_2^2$, true positive rate (TPR) and false positive rate (FPR) of \texttt{CDCD} over 20 simulations. }
\label{tb:tprfpr}
\end{table}

In Table \ref{tb:tprfpr}, we report the average $\ell_2$-estimation error of \texttt{CDCD} for the Cholesky factor measured by $\|\hat\bphi-\bphi\|_2^2$. It is seen that the estimation error of \texttt{CDCD} decreases as $n$ increases and slightly increases as $q$ increases, confirming the results of Theorem \ref{thm2}.
Additionally, in the same table, we report the true positive rate (TPR) and the false positive rate (FPR) of \texttt{CDCD} in terms of identifying the non-zero elements in $\bphi$, or equivalently in $\bT_0, \bT_1,\ldots,\bT_q$. Both TPR and FPR exhibit consistent behavior as the estimation error.
Note that the selection accuracy cannot be fairly evaluated from other methods, as \texttt{DenseSample} and \texttt{SparseSample} do not estimate $\bT_1,\ldots,\bT_q$ and \texttt{SparseCovReg} may produce dense estimators for $\bT_k, k=0,\ldots,q$.

\section{Real Data Analysis}
\label{sec:real}

We apply our proposed \texttt{CDCD} to a co-expression QTLs analysis to recover both the population-level and individual-level covariance matrices of gene expressions. We use the REMBRANDT study (GSE108476) data on 178 patients $(n=178)$ with glioblastoma multiforme (GBM), the most common malignant form of brain tumor in adults \citep{akhavan2010mtor}.
The raw data were pre-processed and normalized using standard pipelines; see \citet{gusev2018rembrandt} for more details.

We fit \texttt{CDCD} to model the expression levels of 73 genes $(p=73)$ that belong to the human glioma pathway in the Kyoto Encyclopedia of Genes and Genomes (KEGG) database \citep{kanehisa2000kegg}. 
For this, we order the 73 genes according to the reference signaling pathway from the KEGG human glioma pathway (\citet{kanehisa2000kegg}; see Supplementary Materials Figure S2). 
As covariates, we consider 118 local SNPs (i.e., SNPs that fall within 2kb upstream and 0.5kb downstream of the gene)
, which are coded with “0” indicating homozygous in the major allele and “1” otherwise, and also include age (continuous) and sex as covariates $(q=120)$. 
We standardize both gene expressions and all covariates to have mean zero and one standard deviation. 
Tuning parameters have been selected by 5-fold cross validation.
We use $\wh{\bT}_k$ and $\hat{\beta}_{t,k}$ for $t \in [p], k=0,\ldots,q$ to denote the estimated $\bT_k$ and $\beta_{t,k}$ by \texttt{CDCD}.

\begin{figure}[!htb]
\centering
\scalebox{0.66}{
\mbox{
			\begin{overpic}[width=3.1in,angle=0]
				{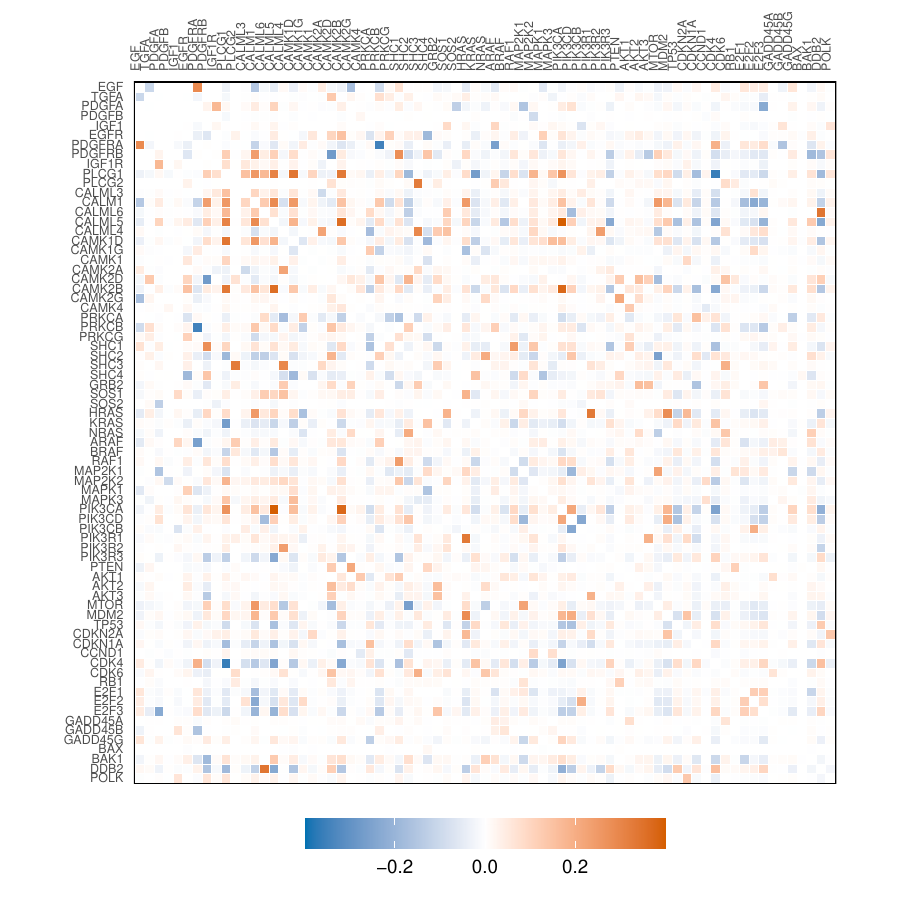}
            \put(10,104){{\uline{\small{\texttt{population-level covariance matrix}}}}}
			\end{overpic}
		
	   }
\mbox{
			\begin{overpic}[width=3.1in,angle=0]
				{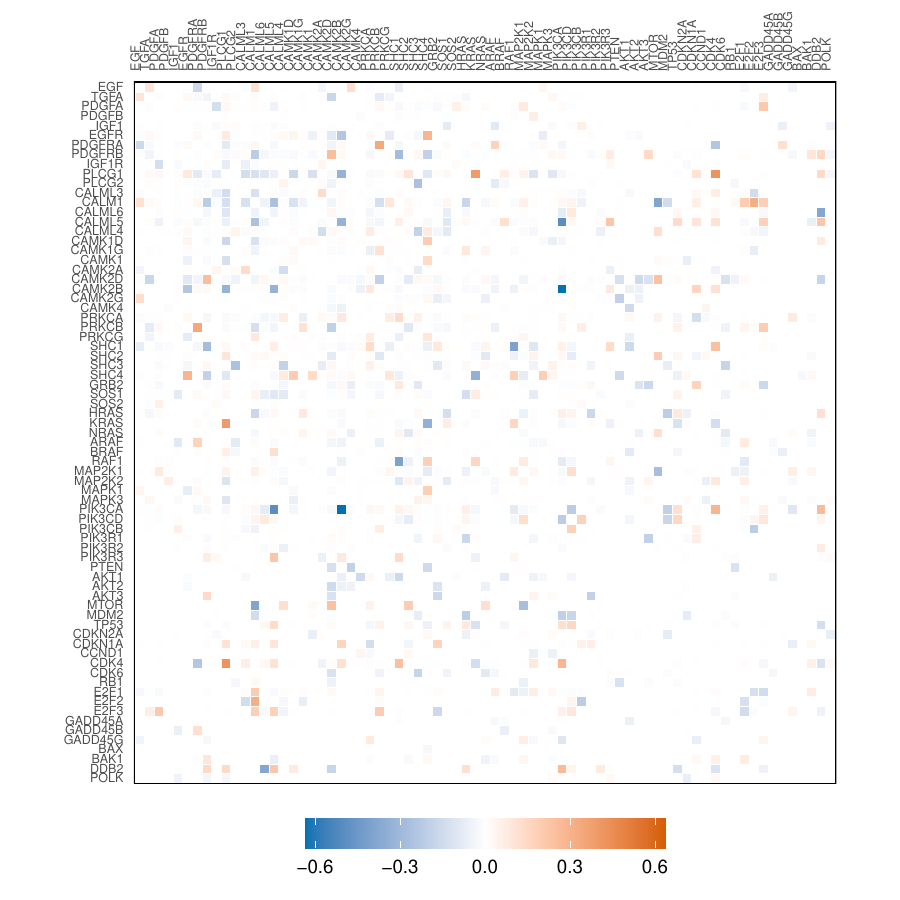}
            \put(10,104){{\uline{\small{\texttt{population-level precision matrix}}}}}
			\end{overpic}
		
	   }
\mbox{
			\begin{overpic}[width=3.1in,angle=0]
				{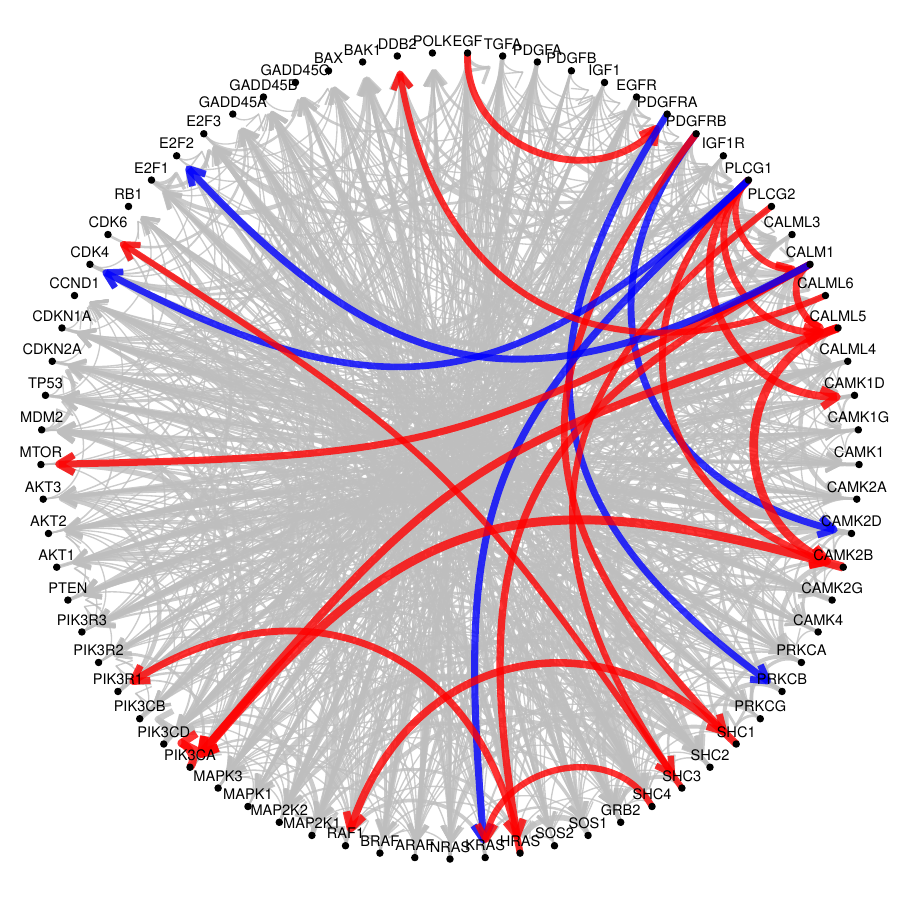}
            \put(2,104){{\uline{\small{\texttt{population-level regression coefficient}}}}}
            \put(88,3){\includegraphics[width=0.33in,angle=0]{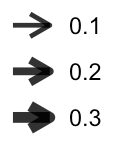}}
			\end{overpic}
		
	   }
}
\caption{Heatmaps of the covariance matrix (left) and the precision matrix (middle) and the linear effects from the independent variables to the dependent variables under the model \eqref{eq:varyingAR} (right) at the population-level. In the heatmaps, the positive elements are shown in red and the negative elements are shown in blue. In the right panel, the width of each arrow is proportional to its effect size and the arrows with effect size above 0.2 are highlighted in red (positive) or blue (negative).}
\label{fig:netpop}
\end{figure}

We first investigate the population-level linear dependence among gene expressions. 
With our proposed \texttt{CDCD}, we can compute the estimated precision matrix and covariance matrix at the population level as $\wh{\bT}_0^\top \wh{\bD}_0^{-1} \wh{\bT}_0$ and its inverse, respectively, where $\wh{\bD}_0$ is a diagonal matrix with $e^{\hat{\beta}_{t,0}}$ as its $t$th diagonal entry.
The estimated covariance matrix at the population level is seen in the left panel of Figure \ref{fig:netpop}. The strong correlation between PIK3CA and genes in the calcium signaling pathway including CALML5, CALM1, CAMK1D, and CAMK2B and the strong negative correlation between CDK4 and PLCG1 were also identified in other works such as \citet{kim2024high}.
The precision matrix at the population level informs on conditional dependence structure among response variables under the Gaussian graphical model, and the estimated precision matrix shown in the middle panel of Figure \ref{fig:netpop} exhibits similar conditional dependence structure to the results of previous literature \citep{zhang2022high, zhang2022multi,meng2024statistical}, including the dependence between DDB2 and CALML6, between PRKCB and PDGFRA and between MTOR and CALM1. 

Furthermore, the dependence structure identified by $\wh{\bT}_0$ has statistical interpretation as regression coefficients under the recursive regression \eqref{eq:varyingAR}. In the right panel of Figure \ref{fig:netpop}, 
the 73 genes are arranged clockwise following the KEGG signaling pathway, with arrows indicating the direction of influence from independent variables to dependent variables under the model \eqref{eq:varyingAR}.
This result not only confirms the linear dependence among genes identified from the covariance or the inverse covariance matrix but also provides insight on the influence from one signaling pathway to another as depicted in the KEGG signaling pathway. For example, the arrow from CALM1 to HRAS represents the influence from the calcium signaling pathway to the Ras-Raf-MEK-ERK pathway, and the arrow from HRAS to PIK3R1 represents the influence from the Ras-Raf-MEK-ERK pathway to the PI3K/ AKT/MTOR pathway, as suggested in the KEGG signaling pathway.

\begin{figure}[!htb]
\centering
\scalebox{0.66}{
\mbox{
			\begin{overpic}[width=3.1in,angle=0]
				{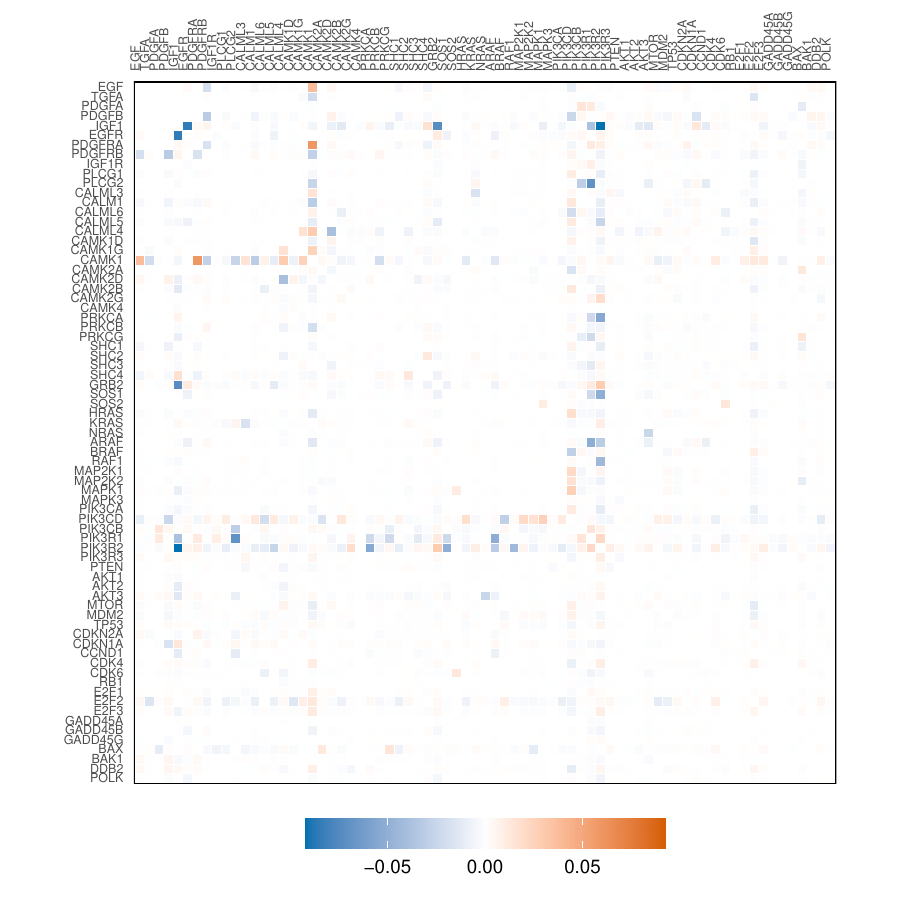}
            \put(18,104){{\uline{\small{\texttt{effect on covariance matrix}}}}}
            \put(-9,42){\rotatebox{90}{{\uline{\texttt{rs6701524}}}}}
			\end{overpic}
		
	   }
\mbox{
			\begin{overpic}[width=3.1in,angle=0]
				{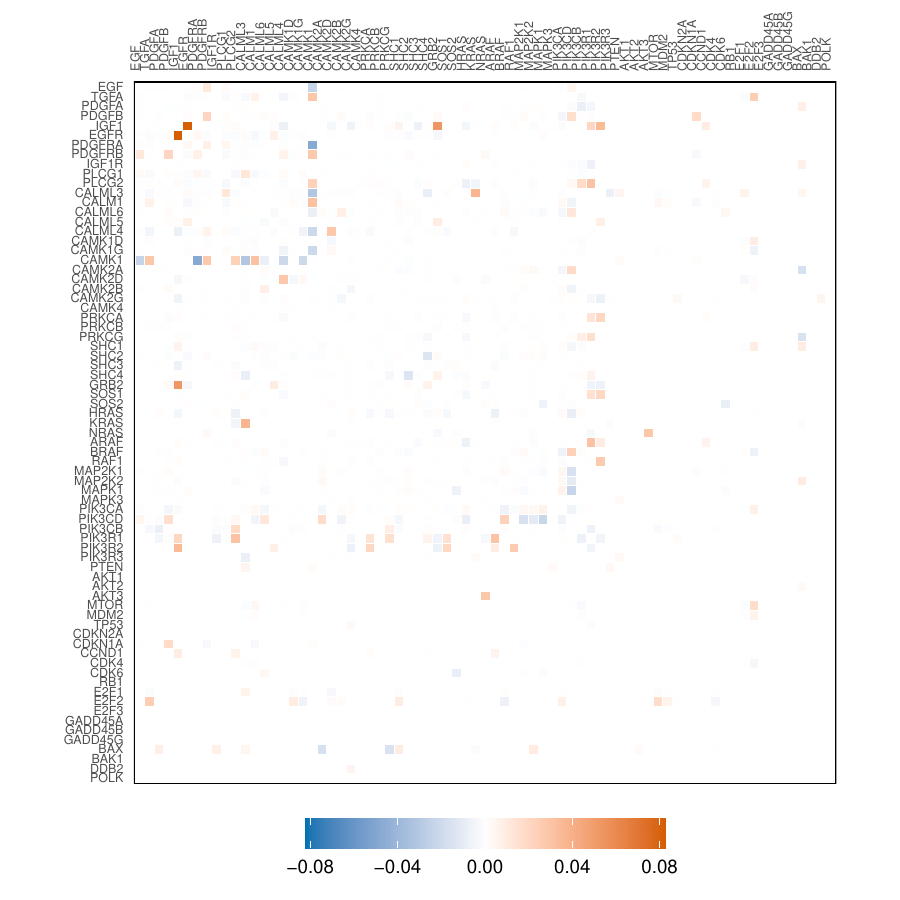}
            \put(18,104){{\uline{\small{\texttt{effect on precision matrix}}}}}
			\end{overpic}
		
	   }
\mbox{
			\begin{overpic}[width=3.1in,angle=0]
				{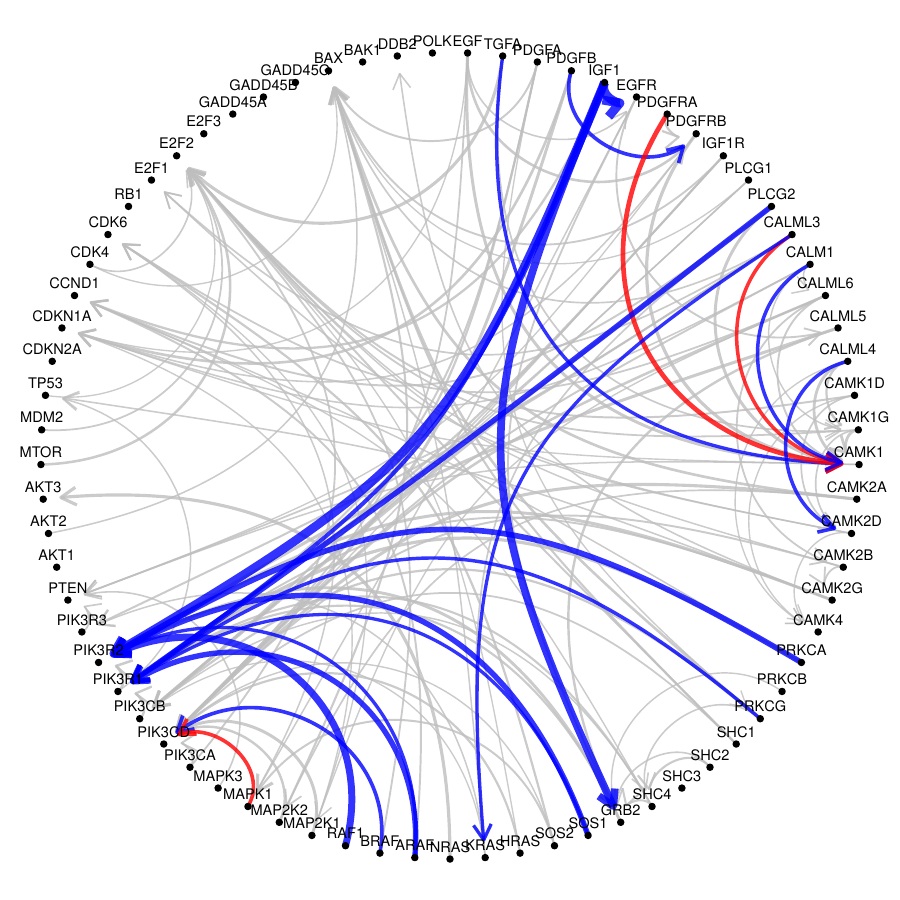}
            \put(9,104){{\uline{\small{\texttt{effect on regression coefficient}}}}}
            \put(88,3){\includegraphics[width=0.4in,angle=0]{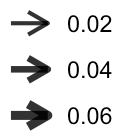}}
			\end{overpic}
		
	   }
}
\vspace*{6mm}
\scalebox{0.66}{
\mbox{
			\begin{overpic}[width=3.1in,angle=0]
				{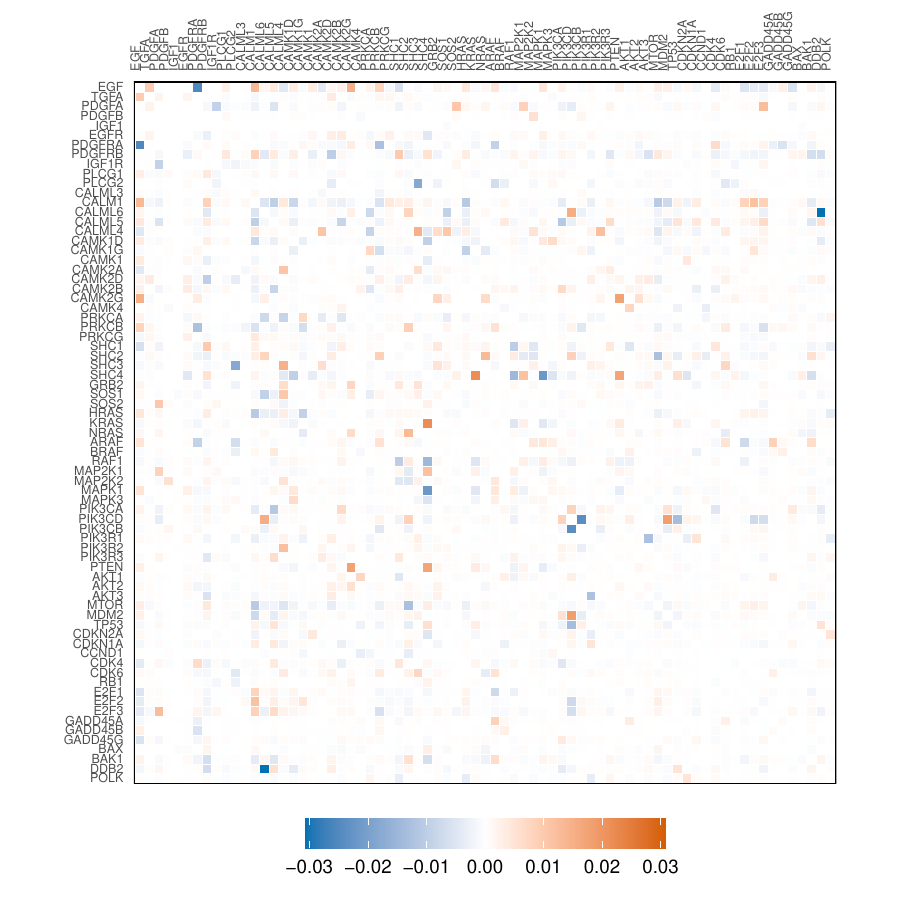}
            \put(-9,42){\rotatebox{90}{{\uline{\texttt{rs9303504}}}}}
			\end{overpic}
		
	   }
\mbox{
			\begin{overpic}[width=3.1in,angle=0]
				{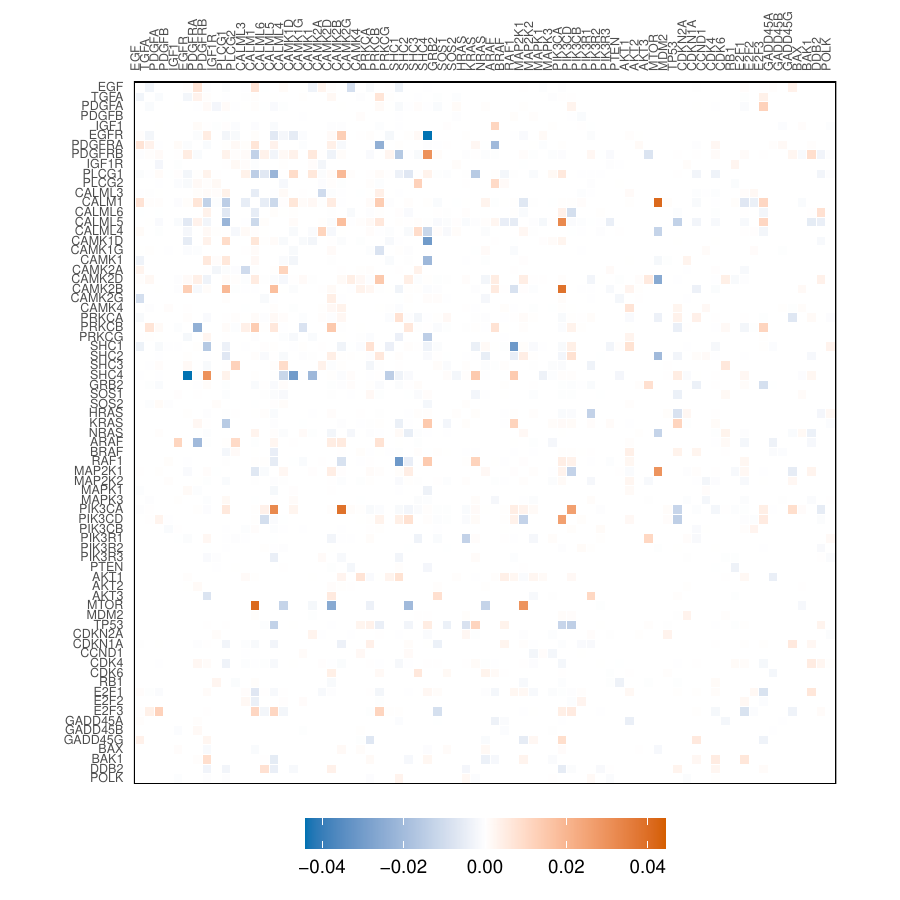}
			\end{overpic}
		
	   }
\mbox{
			\begin{overpic}[width=3.1in,angle=0]
				{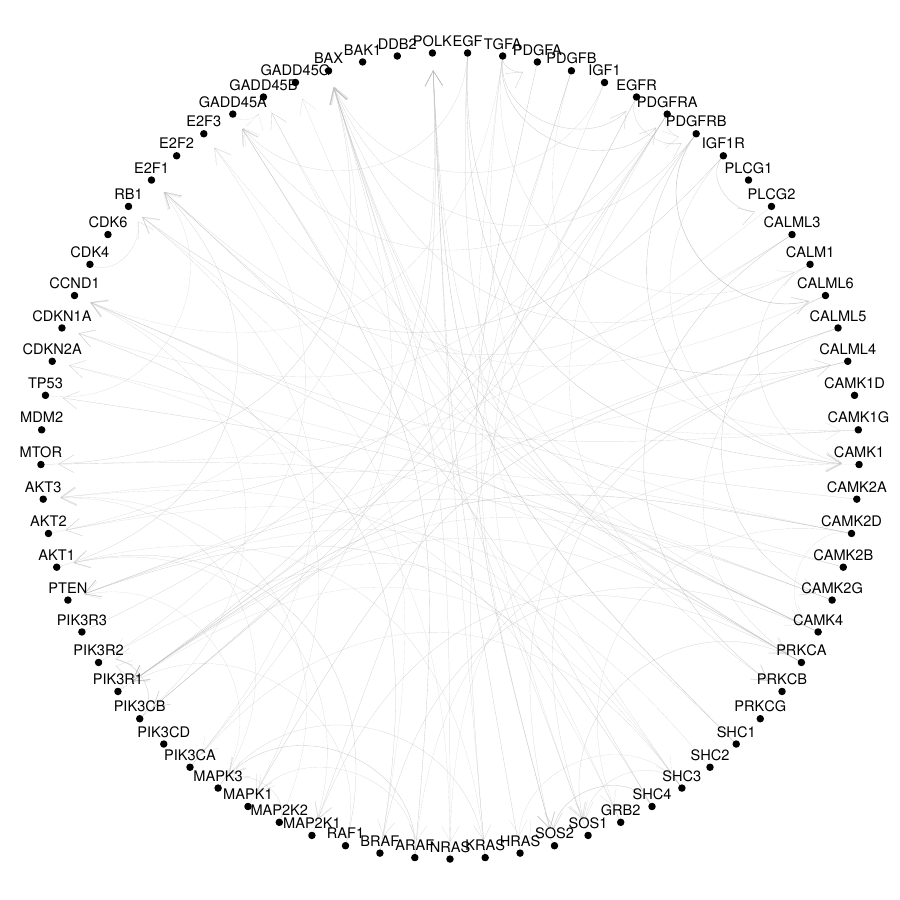}
            \put(86,3){\includegraphics[width=0.4in,angle=0]{figures/legend_net_rs6701524.png}}
			\end{overpic}
		
	   }
}
\caption{Heatmaps of the effects of \texttt{rs6701524} (top) and \texttt{rs9303504} (bottom) to the covariance matrix (left), the precision matrix (middle) and on the linear effects from the independent variables to the dependent variables under the model \eqref{eq:varyingAR} (right). In the heatmaps, the positive elements are shown in red and the negative elements are shown in blue. In the right panel, the width of each arrow is proportional to its effect size and the arrows with effect size above 0.02 are highlighted in red (positive) or blue (negative).}
\label{fig:netcov}
\end{figure}

Next, we examine the covariate effects on the linear dependence among genes. 
For the $k$th binary covariate ($k \in [p]$), we can compute the estimated covariate effect on precision matrix by subtracting the population-level covariance $\wh{\bT}_0^\top \wh{\bD}_0^{-1} \wh{\bT}_0$ from $(\wh{\bT}_0+\wh{\bT}_k)^\top \wh{\bD}_k^{-1} (\wh{\bT}_0+\wh{\bT}_k)$ where $\wh{\bD}_k$ is a diagonal matrix with $e^{\hat{\beta}_{t,0}+\hat{\beta}_{t,k}}$ as its $t$th diagonal entry, and similarly for the covariate effect on covariance matrix.

Non-zero covariate effects have been identified for six SNPs: \texttt{rs6701524}, \texttt{rs10509346}, \texttt{rs10519202}, \texttt{rs1347069}, \texttt{rs9303504}, and \texttt{rs503314}. 
Among them, \texttt{rs6701524}, \texttt{rs10509346}, \texttt{rs1347069} and \texttt{rs503314} have already been identified in previous studies \citep{zhang2022high, zhang2022multi, meng2024statistical, kim2024high} and many of their effects on covariance or precision matrices have already been identified. 
For example, as seen in top panels of Figure \ref{fig:netcov}, the positive effect of \texttt{rs6701524} on covariance matrix between PIK3CD and the genes in the Ras-Raf-MEK-ERK pathway such as MAP2K1, MAP2K2 and HRAS, and the negative effect between PDGFRB and CAMK1 were also detected in \citet{kim2024high}. Also, its negative effect on precision matrix between PIK3R1 and PIK3R2 was selected in \citet{zhang2022multi}, and its positive effect between PIK3R2 and CALML5 was selected in \citet{zhang2022high}.

However, our proposed \texttt{CDCD} not only supports previous findings on the covariate effects on covariance or precision matrices but also allows us to interpret those effects as effects on regression coefficients under \eqref{eq:varyingAR}. For example, in the top right panel of Figure \ref{fig:netcov}, the genes in the Ras-Raf-MEK-ERK pathway are identified as independent variables that affect PIK3CD, PIK3R1 and PIK3R2. This result is a new finding because it suggests that the influence from the Ras-Raf-MEK-ERK pathway to the PI3K/
AKT/MTOR pathway, as depicted in the KEGG signaling pathway, may be strengthened by the effect of \texttt{rs6701524}.
Also, \texttt{CDCD} detects other interesting covariate effects from \texttt{rs6701524}.
For example, co-expressions of PIK3R1 and PIK3R2 with SOS1 are affected by this SNP. This is an interesting finding as SOS1 has been suggested as a potential therapeutic target for GBM \citep{lv2013mir}. Also, PI3K/MTOR is a key pathway in the development and progression of GBM, and the inhibition of PI3K/MTOR signaling was found effective in increasing survival with GBM tumor \citep{batsios2019pi3k}.

Furthermore, we have found some novel co-expression QTLs. For example, as seen in the bottom left and middle panels of Figure \ref{fig:netcov}, the covariance matrix entries within the Ras-Raf-MEK-ERK pathway, particularly between SHC4 and other genes such as KRAS, RAF1, MAP2K1 and MAPK1, are affected by \texttt{rs9303504} and the precision matrix entries between the calcium signaling pathway (e.g. CALM1, CALML5, CAMK2B) and the PI3K/MTOR pathway (e.g. MTOR, PIK3CA) are affected by this SNP.
Interestingly, many of these pronounced entries in the effects on covariance or precision matrices do not appear as arrows in the bottom right panel of Figure \ref{fig:netcov}. 
This is because \texttt{rs9303504} affects the covariance and the precision matrix mainly via $\bD(\bx)$ rather than via $\bT(\bx)$. For example, the largest entry in 
the estimated $\bbeta_{\mydot,k}$
for the effect of \texttt{rs9303504} appear for SHC4, indicating that the variance of this gene expression is affected by this SNP. We have checked that the variance of SHC4 differs by each subgroup: $0.62$ for patients with $\texttt{rs9303504}=0$ and $1.35$ for patients with $\texttt{rs9303504}=1$. These findings were not discovered in previous research on covariate-dependent Gaussian graphical model \citep{zhang2022high, zhang2022multi} because their method assumed the diagonals of the precision matrix are not covariate-dependent.

Co-expressions affected by other SNPs are also worth investigating. For example, \texttt{rs1347069} has been found to affect co-expressions between IGF1R and AKT1, which supports previous findings in \citet{zhang2022multi, meng2024statistical}. \texttt{rs10509346} affects co-expressions between SHC4 and MAPK3, supporting results in \citet{zhang2022high, meng2024statistical}. Also, \texttt{rs503314} was found to influence the co-expressions of CCND1 with PLCG2 and CALML3 by \citet{zhang2022multi} but our method has found that some more genes in the PI3K/MTOR signaling pathway such as PIK3R2, PIK3CA and MTOR are connected to CCND1 by the effect of this SNP. This is interesting because CCND1 has been suggested as altered biomarkers serving as functional targets for personalized treatment of GBM \citep{verdugo2022update}. The effects of other SNPs are summarized in Supplementary Materials S5. 

\section{Discussion}
\label{sec:conc}
Our current method requires a pre-defined ordering of the response variables. One may identify the unknown order of the response variables in data driven approaches such as the  Isomap approach by \citet{wagaman2009discovering} and the Best Permutation Algorithm by \citet{rajaratnam2013best}. 
Our proposed \texttt{CDCD} can be combined with these ordering methods in a two-step approach, that is, we order the response variables in the first step, and then we estimate the covariate-dependent covariance matrix by \texttt{CDCD} in the second step. Furthermore, this approach is useful not only for the covariance estimation but also for learning subject-level directed acyclic graphs (DAG) which is actively being studied in recent literature \citep{ni2019bayesian, sagar2022bayesian, choi2025bayesian, marchant2025covariate}.

On the other hand, instead of identifying a specific order of variables, ensemble models which take the average of multiple permutations have been studied. For example, \citet{zheng2017cholesky} employed the ensemble model via the Cholesky-based model averaging for the covariance estimation, which has extensively been studied further in \citet{kang2020improved, liang2024new, kang2025block, kang2025weighted}. Although this approach requires no prior knowledge on the variable ordering, multiple permutations of the response variables need to be considered, which may be computationally intensive 
and we leave this avenue of research for future enhancement.

\bibliographystyle{asa}
\begingroup
\baselineskip=17.5pt
\bibliography{covlm}
\endgroup

\end{document}